\documentclass{amsart}

\usepackage{amsmath}
\usepackage{amsthm}
\usepackage{amsfonts}
\usepackage{amssymb}
\usepackage{graphicx}
\usepackage{xcolor}
\usepackage{hyperref}
\usepackage{subcaption}
\usepackage{etoolbox}
\usepackage[foot]{amsaddr}
\usepackage{etoolbox}
\usepackage{multicol}
\usepackage{tcolorbox}
\usepackage{datetime}
    \newdateformat{monthyeardate}{%
  \monthname[\THEMONTH], \THEYEAR}
\usepackage{setspace}
\usepackage{longtable}
\usepackage[authoryear,compress]{natbib}

\hypersetup{
    colorlinks,
    linkcolor={red!50!black},
    citecolor={blue!50!black},
    urlcolor={blue!80!black}
}

\makeatletter
    \newtheoremstyle{indented}
        {12pt}% space before
        {12pt}% space after
        {\addtolength{\@totalleftmargin}{3.5em}
        \addtolength{\linewidth}{-3.5em}
        \parshape 1 3.5em \linewidth}% body font
        {}% indent
        {\bfseries}% header font
        {:}% punctuation
            {.5em}% after theorem header
        {}% header specification (empty for default)
\makeatother

\makeatletter
    \newtheoremstyle{indentedProp}
        {12pt}% space before
        {12pt}% space after
        {\addtolength{\@totalleftmargin}{3.5em}
        \addtolength{\linewidth}{-3.5em}
        \parshape 1 3.5em \linewidth}% body font
        {}% indent
        {\bfseries}% header font
        {:}% punctuation
            {0.5em}% after theorem header
        {}% header specification (empty for default)
\makeatother

\theoremstyle{indented}
    \newtheorem{definition}{Definition}[section]
\theoremstyle{indentedProp}
    
\theoremstyle{indented}
    
\theoremstyle{indented}
    
\theoremstyle{indented}
    \newtheorem{theorem}{Theorem}[section]
\theoremstyle{indented}
    
\theoremstyle{indented}

\makeatletter
    \patchcmd{\NAT@test}{\else \NAT@nm}{\else \NAT@nmfmt{\NAT@nm}}{}{}
    \DeclareRobustCommand\citepos
        {\begingroup
        \let\NAT@nmfmt\NAT@posfmt% ...except with a different name format
        \NAT@swafalse\let\NAT@ctype\z@\NAT@partrue
        \@ifstar{\NAT@fulltrue\NAT@citetp}{\NAT@fullfalse\NAT@citetp}}
    \let\NAT@orig@nmfmt\NAT@nmfmt
    \def\NAT@posfmt#1{\NAT@orig@nmfmt{#1's}}
\makeatother

\title[Learning from Learning Machines]{Learning From Learning Machines: \\ Optimisation, Rules, and Social Norms}

\author[Travis LaCroix \& Yoshua Bengio]{Travis LaCroix$^{1,2}$ \\ Yoshua Bengio$^{2,3}$}
\address[$^{1}$]{Department of Logic and Philosophy of Science \\ University of California, Irvine}
\address[$^{2}$]{Mila \\ (Qu{\'e}bec AI Institute / Institut Qu{\'e}b{\'e}cois d'Intelligence Artificielle)}
\address[$^{3}$]{Department of Computer Science \& Operations Research \\ Universit{\'e} de Montr{\'e}al}

\email[Corresponding author]{tlacroix@uci.edu (corresponding author)}
\email{yoshua.bengio@mila.quebec}

\date{Unpublished draft of \monthyeardate\today. {\it Please cite published version, if available}}

\begin{document}

\maketitle

\begin{abstract}
\singlespacing
    There is an analogy between machine learning systems and economic entities in that they are both adaptive, and their behaviour is specified in a more or less explicit way. It appears that the area of AI that is most analogous to the behaviour of economic entities is that of {\it morally good decision-making}, but it is an open question as to how precisely moral behaviour can be achieved in an AI system. This paper explores the analogy between these two complex systems, and we suggest that a clearer understanding of this apparent analogy may help us forward in both the socio-economic domain and the AI domain: known results in economics may help inform feasible solutions in AI safety, but also known results in AI may inform economic policy. If this claim is correct, then the recent successes of deep learning for AI suggest that more {\it implicit} specifications work better than explicit ones for solving such problems. 
    
    \phantom{a}

\end{abstract}

\setcounter{page}{1}

\section{Introduction}
\label{sec:Introduction}

There is an analogy between machine-learning systems and economic entities (such as corporations) in that they are both adaptive, and their behaviour is specified in a more-or-less explicit way. That is, they both seek to optimise something (like profit, rewards, or some other measure of performance), and they often do so under some set of constraints. The pressure that is exerted on these adaptive systems may be specified by precise rules, such as explicit laws and regulations in the economic case, or constraints on parameters and outputs in the case of learning machines---e.g., that output probabilities must be between $0$ and $1$. However, these constraints may also be implemented in terms of more {\it implicit} pressures. For example, an imperative to maximise profit or rewards constrains the action space implicitly, since an entity tries to act in a way that avoids poor outcomes. In this latter case, there are significantly more degrees of freedom that may be allowed, since it is not explicitly specified {\it how} such an objective is achieved.

It would appear that the area of artificial intelligence (AI) that is most analogous to the behaviour of economic entities is that of {\it morally good decision-making}: if an economic or machine-learning system is to achieve its goals (maximising some reward), we want it to do so while acting in a way that is consistent with human social norms and moral principles. Unfortunately, it is
not easy to formally specify a set of rules or a quantifiable objective which
characterises morally good behaviour.

In machine learning, it is thus an open question as to {\it how} precisely moral behaviour can be achieved in an AI system. Part of the problem lies in the fact that, even though humans have tried for a long time, we have not come up with a simple set of programmable rules (constraints) or objectives (such as measurable rewards) with which to provide these systems in the first place. We argue that the resulting mismatch between the behaviour of an entity governed by formal specifications and social norms becomes all the more serious as we consider more powerful entities, such as large multinational corporations or increasingly powerful AI systems. The choices made by these systems may end up being misaligned with society's interests, and the increasing power of these entities could lead to great havoc, on a larger scale, as these technologies or corporate entities are further integrated into society.

Our main claim in this paper is that, by opposition to rule-based systems, the success of {\it deep learning} for AI, along with some of the mathematical hypotheses for this success, suggests that more implicit specifications work better (i.e., are more effective) than explicit ones based on hard rules for representing the kind of complex behaviour needed to align with human morality. We suggest that it is a logical consequence of optimising a reward (like profit) under some set of constraints that leads to border-solutions which touch precisely upon the boundary of one or more of those constraints (like laws)---this consequence is an artefact of the more extensive system in which these entities operate. As such, it should be unsurprising, {\it given that} the goals of these entities may be formalised as an optimisation problem, that they should find solutions that are on the boundary of the permissible action space. In the case of economic entities, constraints are often provided in the form of explicit laws, meaning that it should be unsurprising when such entities converge on solutions which are precisely on the border of what is {\it legal}. This has the further consequence that small perturbations in such a dynamical system or imperfect specification of laws may well lead to genuinely {\it immoral} or otherwise harmful actions, and this would be true both for AI systems and for corporations. Furthermore, as the dimensionality of the action space of these entities increases, we use deep learning theory to argue that it becomes more difficult to approximate a principled objective (like `do no harm') with a set of rule-like constraints since the size of the required set increases (possibly exponentially), along with an increase in opportunities for misspecification and misalignment between the optimised behaviour of the entity and the intended objective.

If the above suggestion is correct---i.e., that there is an analogy between machine-learning systems and economic entities---then our main claim implies that a regulatory and fiscal environment which provides more {\it implicit} pressures on economic entities will lead to their behaviour being more aligned with a common good. Currently, the best-known way to achieve such implicit pressure is via legal or regulatory principles (by opposition to explicit rules) which can be evaluated by human beings, who hold in their head some representation of their version of what is morally acceptable according to the stated principles.

To show that this implication is plausible, this paper has two main objectives. First, we must make explicit the assumptions that underlie our main claim, which we clearly formulate, thus.
\begin{quote}
    {\bf Main Claim}:\\
    {\it The success of deep learning for AI and its theoretical grounding suggests that more implicit specifications are better than explicit ones for representing complex functions of the kind which human brains can represent intuitively but that are not easily reducible to a few formal rules}.
\end{quote}
We will refer back to this claim throughout the paper. Furthermore, when there is a mismatch between the desired morality function and a set of rule-like constraints meant to approximate it, optimising a reward (like profit) under these constraints will lead to borderline solutions which tend to fail the test of morality. To make explicit what would be required for our main claim to be true, we must (1) discuss optimisation problems, (2) explain how deep learning has been more successful than traditional techniques in solving such problems, and (3) clarify how constraints are implicit in this domain. This is the goal of Section~\ref{sec:Assumptions}. In Section~\ref{sec:Assumption1}, we begin our analysis by offering a brief overview of the mechanics of optimisation problems, which further provides some technical background for our  audience outside of the machine-learning community. In Section~\ref{sec:Assumption2}, we make clear in what ways deep learning {\it is} successful as compared with traditional machine-learning techniques. In Section~\ref{sec:Assumption3}, we discuss the distinction between explicit constraints in traditional methods and implicit constraints in deep learning.

Once the truth of our main claim has been established, we draw the following conclusion: given the close analogy between machine-learning systems and economic entities, more implicit specifications (or specifications that are more implicit)
and less explicit rules (or fewer rules that are explicit) may also be more successful for economic entities to achieve their goals. We will refer to this as our {\it hypothetical analogy}. To assert that the hypothetical analogy is plausible, it is necessary to clarify the similarity between machine-learning systems and economic entities. This is the goal of Section~\ref{sec:Analogies}. This implication takes the form of an {\it analogical argument}. Therefore, in Section~\ref{sec:AnalogicalArgument}, we discuss analogies in general, and what it takes for an analogical argument to be strong. In Section~\ref{sec:ForwardDirection}, we review some literature from economics in the forward direction of our analogy; it is well-accepted that solutions in economics can help to inform machine learning. The principal contribution of our paper is given in Section~\ref{sec:BackwardDirection}, where we discuss the backward direction of the analogy---namely, that results from machine learning may inform policies in economics and social organisation in general.

The main theoretical result of this analogy is that implicit specifications may be more well-suited for finding solutions which are both optimal {\it and} beneficial. However, this has the immediate consequence that we must give a clear statement of {\it implicit specifications} to move toward practical progress in this domain. The plausibility of this is discussed in Section~\ref{sec:Possibility}. Finally, in Section~\ref{sec:Conclusion}, we offer some examples of how this view may proceed in terms of implementation, thus paving the way for future research to build upon this framework.

\section{Assumptions}
\label{sec:Assumptions}

In this section, we (1) discuss optimisation problems, (2) explain how deep learning has been more successful than traditional techniques in solving such problems and explicate some of the relevant theoretical 
analyses, and (3) clarify how constraints are implicit in the deep-learning domain. Making these three things explicit should be sufficient for the truth of our main claim: that the success of deep learning for AI suggests that more implicit specifications are better than explicit ones for solving optimisation problems under the implicit constraint of acting morally.

\subsection{On Optimisation Problems}
\label{sec:Assumption1}

The vast majority of machine-learning algorithms train models and perform inference by solving {\it optimisation problems} \citep{MurphyBook2012,Goodfellow-et-al-2016}. In mathematics and computer science, an optimisation problem involves finding the {\it best} solution from the set of all ({\it feasible}) solutions. More specifically, given a metric, $\mathbf{P}$, the solution to an optimisation problem consists in finding a function (or the parameters of a function) within a given set (of feasible or acceptable solutions) that maximises (minimises) this metric. In machine learning, this metric is usually an average on some data or an expectation over a distribution. Optimisation problems are ubiquitous in applied mathematics, and methods for solving such problems are essential for many different fields, including manufacturing and production, inventory control, transportation, scheduling, finance, marketing, engineering and control, mechanics, economics, and policy modelling, to name a few~\citep{NumOptBook}. 

A typical optimisation problem consists of an {\it objective function}, $R(x)$, which gives the quantity to be maximised (e.g. a reward); a set of {\it variables}, $\{ x_1, x_2, x_3, \ldots\} = \{ x_n \}$, which can be set by an agent and constitute the inputs for the objective function; and a set of {\it equality constraints}, $\{ h_n(x) \}$, or {\it inequality constraints}, $\{ g_n(x)\}$, which limit the possible values for the variables.\footnote{Of course, this is a gross simplification: some problems may not have constraints; there may be one or many variables, which are discrete or continuous; optimisation problems themselves may be static or dynamic, equations may be linear or nonlinear, and systems may be deterministic or stochastic.} 

{\it Convex} optimisation problems are standard in computer science, and many classes of convex optimisation problems admit polynomial-time algorithms \citep{Nesterov-Nemirovskii-1994}, though mathematical optimisation generally is NP-hard \citep{Sahni-1974, Murty-Kabadi-1987, Pardalos-Vavasis-1991}.\footnote{In complexity theory, a decision problem can be understood as a problem with a {\it yes} or {\it no} answer. {\bf P} is the complexity class denoting the set of all such decision problems that can be decided in polynomial time; for example, many mathematical operations, such as addition and subtraction, can be computed in polynomial time. {\bf NP} is the class denoting the set of problems where proofs of a positive solution (a `yes' answer to the decision problem) can be {\it verified} in polynomial time. {\bf NP-Hard} is the class of decision problems that are as difficult as any {\bf NP} problem to solve. For example, the {\it Halting Problem} is NP-Hard.} Indeed, while many non-convex  optimisation problems are NP-hard to solve {\it optimally}, there is a range of such problems that are known to be NP-hard to solve {\it approximately}, as well \citep{Meka-et-al-2008}. The standard form of a convex optimisation problem can be written as follows:
\begin{equation}\label{eq:ConvOpt}
    \begin{array}{rrcll}
        \underset{\mathbf{x}}{\mathrm{maximise}} & R({\mathbf x}) & & & \\
        & & & & \\
        \rm{subject\ to} & g_{i}({\mathbf x}) & \leq & 0, & i = 1, \ldots m \\
        & h_{i}({\mathbf x}) & = & 0, & i = 1, \ldots p \\
    \end{array}
\end{equation}
Here, $\mathbf{x} \in \mathbb{R}^{n}$ is the vector of optimisation variables, the functions $R, g_1 , \ldots , g_m$ are convex, and the functions $h_1, \ldots, h_p$ are affine. The {\it feasible set} for $R$ is the set of points $\mathbf{x} \in \mathbb{R}^{n}$ satisfying all of the constraints, $g_1 \leq 0, \ldots , g_m \leq 0, h_1 = 0, \ldots , h_p = 0$ \citep{Boyd-Vandenberghe-2004}. Importantly, the optimal solutions of an optimisation problem can occur at every point of the intersection between the domain of the objective function and the set of constraints on the cost function. 

Two important theorems in linear programming are given as follows:
\begin{theorem}
    For a polyhedron $P$ and a point $x\in P$, the following are equivalent:
    \begin{enumerate}
        \item $x$ is a {\it vertex} of $P$,
        \item $x$ is an {\it extreme point} of $P$, and
        \item $x$ is a {\it basic feasible solution}.
    \end{enumerate}
\end{theorem}
\begin{theorem}
    Any bounded LP in standard form has an optimum at a basic feasible solution.
\end{theorem}
Therefore, it follows straightforwardly for this case that it is a logical consequence of placing explicit constraints on (this particular class of) optimisation problems, that solutions will live on the boundary of those constraints. Though this is not true in the general case (e.g., non-convex optimisation problems may have solutions in the interior of the feasible set), the point being made here is still illustrative because (1) some approaches to non-convex optimisation use special formulations of linear programming, and (2) many non-convex optimisation are solved {\it in practice} by reducing the problem into a convex optimisation problem and then finding a solution for that problem (which then approximates the solution to the original problem).

In the economic case, the constraints are given by society in the form of regulations, laws, etc. We have illustrated, via formal consequences, that placing explicit constraints on an action space implies that an optimising agent will find a solution on the boundary of those constraints. The same conclusion can be shown intuitively by the following thought experiment. Consider a typical profit-maximising corporation whose behaviour is explicitly constrained by laws and regulations. Assume further that this corporation is acting optimally, in the sense of maximising its profit. Is it possible for such an entity to increase its profit by removing some of those constraints? If the answer to this question is `yes', then this implies that the  corporation's optimal profit-maximising solution lived on the border of at least one of the (legal or regulatory) constraints imposed on its action space. Furthermore, the lobbying efforts of larger corporations to relax or remove particular laws or regulations provide strong evidence that the answer to this question {\it is}, in fact, `yes'.

\subsection{The Successes of Deep Learning}
\label{sec:Assumption2}

Machine learning is a branch of artificial intelligence that primarily involves designing algorithms which can be used to learn from data or interactions with their environment. A standard definition of machine learning is given by \citet{Mitchell-1997}:
\begin{quote}
    a computer program is said to learn from experience $E$ with respect to some class of tasks $T$ and performance measure $P$ if its performance at tasks in $T$, as measured by $P$, improves with experience $E$. (2)
\end{quote}
In essence, a machine-learning algorithm builds a mathematical model based on structured {\it sample} ({\it training}) data, to make predictions or decisions in novel contexts, without being explicitly programmed to perform that task \citep{Koza-et-al-1996, Bishop-2006}. This is often achieved by approximate optimisation of an objective function, which typically measures how well the learner performs on a set of data or experiences.

Deep learning is a subset of machine learning whose algorithms are loosely inspired by brains (both neuroscience and cognitive science), where artificial neural networks perform the computations. A distinguishing feature of deep learning is its emphasis on learning internal representations, thus relying less on hand-crafted features or attributes. This makes possible the ability to learn from raw data, like images or sounds: `[d]eep learning algorithms seek to exploit the unknown structure in the input distribution in order to discover good representations, often at multiple levels, with higher-level learned features defined in terms of lower-level features' \citep[p. 17]{Bengio-2012}.

Deep learning, as a form of machine learning, has seen great successes in the last decade. Salient examples of this are in {\it computer vision}, {\it speech recognition},  {\it machine translation} and {\it playing games} (and beating the world champions at games like Go). Deep learning's successes can be attributed to its ability to capture complex functions, which are not generally easy to represent as a small set of rules. This ability captures a form of {\it intuitive knowledge} which humans typically have but can hardly express verbally, much less program in computers. This is in contrast with earlier approaches to artificial intelligence (such as expert systems and rule-based systems) which relied on formalising knowledge expressed by humans. These earlier approaches work well when the domain of knowledge considered can be well described in a set of clear and specific rules which one can thus program into a computer. 

\citet{Lewis-Kraus-2016} highlights the progress brought by deep learning in machine translation with the following illustrative example. The opening of Ernest Hemingway's short story, `The Snows of Kilimanjaro' proceeds thus:
\begin{quote}
{\it Original English Text}:

    \noindent Kilimanjaro is a snow-covered mountain 19,710 feet high, and is said to be the highest mountain in Africa.\ldots Close to the western summit there is the dried and frozen carcass of a leopard. No one has explained what the leopard was seeking at that altitude.
\end{quote}
Translating the Japanese version of this text back into English using Google Translate's statistical machine-translation methods, we obtain the following.
\begin{quote}
(A) {\it Machine Translation (Pre-Deep Learning)}:

    \noindent Kilimanjaro is 19,710 feet of the mountain covered with snow, and it is said that the highest mountain in Africa.\ldots The top close to the west, there is a dry, frozen carcass of a leopard. Whether the leopard had what the demand at that altitude, there is no that nobody explained.
\end{quote}
However, after Google rolled out its deep-learning methods, the output is as follows.
\begin{quote}
(B) {\it Machine Translation (Post-Deep Learning)}:

    \noindent Kilimanjaro is a mountain of 19,710 feet covered with snow and is said to be the highest mountain in Africa.\ldots Near the top of the west there is a dry and frozen dead body of leopard. No one has ever explained what leopard wanted at that altitude.
\end{quote}
\citet{Lewis-Kraus-2016} highlights that save for a missing article in the deep-learning translation, it is almost indistinguishable from a human translation. Though it took approximately 13 months to implement, from the user's point of view, the change from (A) to (B) happened overnight.

In addition to this particular example, deep learning has seen great success over traditional techniques in many applications, including computer vision \citep{OMahony-et-al-2019}, speech/audio/image recognition \citep{Lee-et-al-2009, Ngiam-et-al-2011, Mroueh-et-al-2015, Wu-et-al-2015, Amodei-et-al-2016, Prasad-et-al-2020}, natural language processing \citep{Collobert-Weston-2008, Young-et-al-2018}, social network filtering \citep{Xie-et-al-2018} and spam detection \citep{Zheng-et-al-2015, Wu-et-al-2017, Barushka-Hajek-2018}, machine translation \citep{Wu-et-al-2016, Vaswani-et-al-2018}, bioinformatics \citep{Chen-et-al-2016, Min-et-al-2017, Li-et-al-2019}, drug design \citep{Gawehn-et-al-2016, Jing-et-al-2018}, medical image analysis \citep{Litjens-et-al-2017, Shen-et-al-2017}, game playing \citep{Mnih-et-al-2013, Mnih-et-al-2015, Silver-et-al-2016, Silver-et-al-2017, Silver-et-al-2018}, etc. In these and many other fields of application, deep-learning algorithms produce results that are substantially better than previous state-of-the-art, sometimes approaching or surpassing human performance.

\subsection{Explicit and Implicit Knowledge}
\label{sec:Assumption3}

Many of the best state-of-the-art neural network architectures rely on a particular form of nonlinearity in each of the artificial neurons: piecewise-linearity (i.e. formed by the concatenation of linear pieces). The use of these rectifying nonlinearities was inspired by neuroscience and shown experimentally to make it easier to train deeper neural networks \citep{Glorot+al-AI-2011-small}. An interesting theoretical feature of these networks makes their analysis easier: the overall function (from inputs to outputs) computed by such deep networks is piecewise-linear. This means that there exists a set of linear constraints which can be combined logically to represent precisely the same function as that computed by one of these deep networks.

However, it has been demonstrated that the number of such linear pieces (i.e., the rules that would obtain if we were to translate the network into a set of rules) grows exponentially both with the number of artificial neurons per layer and with the number of layers~\citep{Montufar-et-al-NIPS2014}. In other words, deep neural networks cannot be well represented by a reasonably small set of rules. This is consistent with the notion that complex, intuitive knowledge is not easily expressed in a small set of verbally expressible rules (i.e., where each rule involves a simple relationship between a few variables). Whereas a deep neural network thus captures implicit knowledge (difficult or impossible to express explicitly via a small set of rules), classical symbolic AI approaches relied on humans expressing explicit knowledge, which can be formalised into a set of rules.

What is strongly suggested by these theoretical results, in conjunction with the empirical success of deep learning on tasks involving implicit and intuitive knowledge, is the following: trying to capture intuitive and implicit knowledge in an explicit way (the rule-based approach) runs the risk of a substantial gap between that implicit knowledge and the function represented by rule-like constraints. The `size' of that gap can be measured by the remarkable improvement which has been achieved when going from rule-based AI to deep learning on a broad set of tasks which precisely resist straightforward formalisation (like computer vision, speech recognition, machine translation, or playing the game of Go).

Having established this foundation, we move on to fleshing out the analogy between machine-learning systems and economic entities in the next section.

\section{Analogies}
\label{sec:Analogies}

The hypothetical analogy---implicit specifications may also be more successful for economic entities to achieve their goals---is a consequence of our main claim (elucidated in the previous section) and the fact that machine-learning systems and economic systems are analogous. This implication takes the form of an {\it analogical argument}. Therefore, to establish the strength of our hypothetical analogy, we must specify how machine-learning systems and economic entities are analogous. In this section, we begin by describing the formal components of analogies and analogical reasoning (Section~\ref{sec:AnalogicalArgument}), before providing some evidence for the strength of this analogy (Section~\ref{sec:ForwardDirection}), and finally drawing our main conclusions (Section~\ref{sec:BackwardDirection}).

\subsection{The Structure of an Analogical Argument}
\label{sec:AnalogicalArgument}

An analogy is a comparison of the apparent similarity between two objects or systems of objects. Analogical reasoning is a form of reasoning that takes advantage of such apparent similarity. Following \citet{Hesse-1966}, we can distinguish between {\it horizontal} and {\it vertical} relations in an analogy. Horizontal relations are similarity (and difference) relations between domains, whereas vertical relations involve correspondence between objects, properties, and relations within a domain. Following \citet{Keynes-1921}, we can further distinguish between a {\it positive} analogy and a {\it negative} analogy: 
\begin{definition}\label{def:PositiveAnalogy}
{\it Positive Analogy}\\
Let $P = \{ P_1, \ldots , P_n \}$ be a set [or `list'] of accepted propositions about the source domain, $S$. Let $P^{*} = \{ P_{1}^{*}, \ldots , P_{n}^{*} \}$ be a set of corresponding propositions, which are all accepted as holding of the target domain, $T$. $P$ and $P^{*}$ represent accepted (or known) similarities. We refer to $P$ as the {\it positive analogy}.
\end{definition}
\begin{definition}\label{def:NegativeAnalogy}
{\it Negative Analogy}\\
Let $A = \{ A_1, \ldots, A_r \}$ be a set [or `list'] of propositions that are accepted as holding in $S$, and let $B^{*} = \{ B_{1}^{*}, \ldots , B_{s}^{*} \}$ be a set of propositions holding in $T$. Suppose the analogous propositions $A^{*} = \{ A_{1}^{*} , \ldots , A_{n}^{*} \}$ fail to hold in $T$, and the propositions $B = \{ B_1 , \ldots , B_n  \}$ fail to hold in $S$. We can write $A, \neg A^{*}$ and $\neg B, B^{*}$ to represent accepted or known differences and refer to $A$ and $B$ as the {\it negative analogy}.
\end{definition}
\begin{definition}\label{def:NeutralAnalogy}
{\it Neutral Analogy}\\
The {\it neutral analogy} consists of accepted propositions about $S$ for which it is not known whether an analogue, $Q^{*}$, holds in $T$.
\end{definition}
\begin{definition}\label{def:HypotheticalAnalogy}
{\it Hypothetical Analogy}\\
The {\it hypothetical analogy} is the proposition, $Q^{*}$, in the neutral analogy that is the focus of our attention.
\end{definition}

A standard way to understand an analogy is as a mapping of structures \citep{Gentner-Gentner-1983}.\footnote{Technically, a {\it model-theoretic isomorphism}.} On their view, analogies take advantage of `certain aspects of existing knowledge, and that this selected knowledge can be structurally characterized' (p. 101).\footnote{See also \citet{Gentner-1983, Gick-Holyoak-1983, Holyoak-Thagard-1989, Holyoak-Thagard-1995, Forbus-et-al-1994, Forbus-et-al-1995, Forbus-2001, Gentner-et-al-2001b, Dunbar-2001}.} A comparison of two complex concepts by analogy takes advantage of relations between the constituent parts of the complex concept, but does not require that any two objects in that domain are similar. Good analogies are supposed to be characterised, on this view, by {\it systematic} relational correspondences: these are `[a]nalogies are about relations, rather than simple features. No matter what kind of knowledge (causal models, plans, stories, etc.), it is the structural properties (i.e., the interrelationships between the facts) that determine the content of an analogy' \citep[p. 3]{Falkenhainer-et-al-1989}.\footnote{Though, see \citet{Schlimm-2008}.}

An {\it analogical argument} is an explicit representation of analogical reasoning, which depends upon cited similarities between the objects or systems in question, and which supports the (explicit) conclusion that some further similarity exists. An analogical argument is inductive to the extent that the analogy, upon which the argument depends, makes the argument's conclusion plausible (in the sense of enhancing its probability).

In analogical reasoning, there is a noted similarity between a {\it source domain} and a {\it target domain}. The analogy comes into play when noting some property or set of properties, $Q$ (the neutral analogy) in the source and reasoning that the property, $Q$, or some similar property, $Q^{*}$ (the hypothetical analogy), holds in the target \citep{Keynes-1921}. Therefore, given that the source and target contexts share some relevant structural properties, it is plausible that the target context also contains some other property which obtains in the source. \citet{sep-reasoning-analogy} enumerates the following general `commonsense' guidelines for the strength of analogical reasoning:\footnote{See also \citet{Mill-1843, Keynes-1921, Robinson-1930, Stebbing-1933, Moore-Parker-1998, Woods-Irvine-Walton-2004, Copi-Cohen-2005}. Alternative guidelines are given in \citet{Aristotle-Prior-Analytics, Aristotle-Rhetoric, Aristotle-Topics} and \citet{Hesse-1966}. See also \citet{Hume-1779, Quine-Ullian-1970}.}
\begin{enumerate}
    \item The more similarities (between two domains), the stronger the analogy; similarly, the more differences, the weaker the analogy.
    \item The more differences, the weaker the analogy.
    \item The greater the extent of our ignorance about the two domains, the weaker the analogy.
    \item The weaker the conclusion, the more plausible the analogy.
    \item Analogies involving causal relations are more plausible than those not involving causal relations.
    \item Structural analogies are stronger than those based on superficial similarities.
    \item The relevance of the similarities and differences to the conclusion (i.e., to the hypothetical analogy) must be taken into account.
    \item Multiple analogies supporting the same conclusion make the argument stronger.
\end{enumerate}

\citet{Bartha-2010} notes that an analogical {\it argument} has the form: `[i]t is plausible that $Q^*$ holds in the target because of certain known (or accepted) similarities with the source domain, despite certain known (or accepted) differences' (15). Here, plausibility might be simply interpreted as {\it prima facie} plausibility. Furthermore, to say an hypothesis is {\it prima facie} plausible is to say only that (1) it has epistemic support, and (2) it has pragmatic importance. The first of these simply requires an appreciable likelihood of being true (or {\it successful}).

We said in the introduction that there is an analogy between machine-learning systems and economic entities in that they are both adaptive, and their behaviour is specified in a more-or-less explicit way. With this formal framework laid out, we are now in a position to assess the strength of this analogy. If the analogy holds and is strong, then a clearer understanding of it may help us forward, both in the socio-economic domain and in the AI domain. That is, 
\begin{enumerate}
    \item[$\Rightarrow$] Known results in economics may help inform feasible solutions in AI safety; but also,
    \item[$\Leftarrow$] known results in AI may inform economic policy (with respect to, e.g., regulation).
\end{enumerate}
The next section (\ref{sec:ForwardDirection}) discusses the forward direction by appealing to work in economics that has suggested places where the socio-economic domain may help to inform solutions in AI safety. 

\subsection{The Forward Direction}
\label{sec:ForwardDirection}

Here, we consider current work in the Economics literature on the forward direction of our analogy. This should give some {\it prima facie} plausibility to the strength of the analogy, thus making the backward direction (and in particular, our hypothetical analogy) more plausible. 

To design an AI agent, we must first specify what we want it to do. In reinforcement learning, for example, this consists in defining a {\it reward} function which tells the agent the value of all state and action combinations. When an algorithm does well at learning, this means that it performs well {\it with respect to} this reward function. In some cases, specifying a reward function may be relatively straightforward---e.g., a win in a game of chess consists of a $+1$ reward, a loss $-1$, and a draw $0$. 

However, in real-world situations, it is often unclear how the reward function should be specified. This leads to the {\it alignment problem}: an artificial system is aligned (i.e., with human goals) just in case it reliably or accurately implements human values. This involves, on the design end, ensuring that the objective given to the AI system is an accurate representation of the intended objective. Alignment becomes a {\it problem} when there are differences between the specified reward function and what relevant features/actions human actually value. The alignment problem comes in many different forms and may include, e.g., negative (unintended) side-effects, reward hacking, limited capacity for human oversight, differences between training and deployment environments, uncontrolled or unexpected exploration after deployment. Misalignment may arise because of the fundamental difficulty in representing and implementing human values---problems of fairness and bias in machine learning algorithms are fundamentally problems of alignment. 

\citet{Hadfield-Menell-Hadfield-2019} suggest that the AI alignment problem has a clear analogue in the human principle-agent problem, which has been long-studied by economists and legal scholars. In the principal-agent problem, a human agent is tasked with taking actions that achieve a principal's objective. Ideally, alignment is achieved through what is called a {\it complete contingent contract}. Such a contract perfectly aligns the agent's incentives with the principal's objective in a a theoretically enforceable way (by, e.g., monetary transfers or punishments, social sanctions, etc.), and which specifies the reward received by the agent for all possible actions and states of the world---i.e., such a contract altogether accounts for every possible contingency.

Economists initially modelled this problem by assuming that the principal agent could simply fine-tune the agent's incentives on everything that she cared about---this is quite analogous to the machine learning problem of designing objective functions to accommodate everything that we care about.\footnote{See discussion in \citet{Russell-2019}.} However, in practice, it turns out to be challenging to design reward functions for many tasks which we would like AI agents to perform. This difficulty also arises in the economic sphere: contracts constructed in this way end up being complex and strange \citep[p. 4]{Hadfield-2019}. 

Furthermore, writing such a contract is routinely impossible since states of the world may be non-contractible (because they are unobservable or unverifiable and because the number of possible states of the world is enormous for all but trivial domains. Further still, human agents are boundedly rational, so they may not be able to evaluate the optimal actions in all states of the world, and they may not even be able to predict or imagine all the possible states of the world. (Similarly, it may be impossible to unambiguously describe every contingency.) Even if these problems are ignored, it may be too costly (either financially, temporally, or computationally) to write every contingency down in a way that can also be enforced at a reasonable cost. Therefore, \citet{Hadfield-Menell-Hadfield-2019} highlight that contracts in human relationships are usually, if not necessarily, incomplete.\footnote{See also \citet{Simon-1982, Grossman-Hart-1986, Williamson-1996, Bernheim-Whinston-1998, Graebner-2009}.} It follows that not all of the expectations arising from a formal contract will be documented. As a result, it is crucial to understand the level of, e.g., detail \citep{Mayer-Argyres-2004} or complexity \citep{Reuer-Arino-2007} that a contract needs to employ to be effective. 

But this is precisely the alignment problem in artificial intelligence systems. The alignment problem is analogous to issues in economics and law surrounding {\it incomplete contracts} \citep{Hadfield-Menell-Hadfield-2019}. It will be impossible for an employer to be able to incentivise their (new) employee on every dimension of a job that they care about; so, one must accept that in delegating a task to another agent, it may not be done exactly in the way that one wants. \citet{Hadfield-2019} suggests that the principle-agent problem underlies everything from employment to democracy. Similarly, we design AI systems to perform tasks of interest, but we want to ensure that the solutions that they do find are ones that are aligned with our own goals and values. Thus, we must design these systems in a way that avoids unwanted or unanticipated solutions, without epistemic access to what those solutions might be in advance. 

Incomplete contracts are analogous to (potentially) misspecified rewards in the following sense. The former involve bounded rationality, costly cognition/drafting, possible non-contractibility, planned renegotiation, planned completion by a third party in the event of a dispute, and (negative) strategic behaviour on the part of the agent. The latter has parallel features, including bounded rationality, costly engineering/design, possible non-implementability, planned iteration on reward functions, expected completion by a third party, and adversarial blinding or reward preservation.

Contracts are often used to align expectations \citep{Macaulay-1963, Argyres-et-al-2007}; however, contracts are also often violated \citep{Robinson-Rousseau-1994}. The two dominant theoretical perspectives on the nature of contracts understand contracts as instruments of control \citep{Macneil-1977, Williamson-1985, Williamson-1991} and instruments of coordination \citep{Argyres-et-al-2007, Mayer-Argyres-2004}. In light of this, \citet{Hadfield-Menell-Hadfield-2019} suggest that 
\begin{quote}
    AI designers, like contract designers, are faced with the challenge of achieving intended goals in light of the limitations that arise from translating those goals into implementable structures to guide agent behavior (learning algorithms and reward functions). An AI is misaligned whenever it chooses behaviors based on a reward function that is different from the true welfare of relevant humans. (p. 3)
\end{quote}
\citet{Hadfield-Menell-Hadfield-2019} posit that an analysis of incomplete contracting in economics and law provides a useful framework for discussing the alignment problem in an AI context---extant technical results in formal economics may provide a way forward for AI researchers. Importantly, however, this `is supported by substantial amounts of external structure, such as generally available institutions (culture, law) that can supply implied terms to fill the gaps in incomplete contracts' (p. 2).  

Thus, \citet{Hadfield-Menell-Hadfield-2019} take the alignment problem for AI and incomplete contracts as analogous to argue for the hypothetical analogy that solving the alignment problem will inevitably require building formal tools that allow AI systems to take into account the costs associated with taking actions deemed wrongful by human communities when they evaluate their rewards---i.e., this will require figuring the {\it normative} structures of human communities. This is drawn from the neutral analogy of incomplete contracting in economic systems. 

In the next section, we look toward the opposite direction, making explicit the positive, neutral, and hypothetical analogies between AI systems and economic entities more generally, and assessing explicitly the analogical argument that we are putting forward.

\subsection{The Backward Direction (The Hypothetical Analogy)}
\label{sec:BackwardDirection}

We suggested in the introduction that there is an analogy between economic entities and machine learning entities and that the most analogous feature of these separate systems is in the realm of morally good decision making. We make the positive and hypothetical analogies between these two systems explicit in Table~\ref{tab:PosAnalogy}.

\begin{table}[htb!]
    \footnotesize 
    \begin{multicols}{2}
        {\bf Machine Learning Systems}
    
        ({\it Source Domain})
    
        \columnbreak
    
        {\bf Economic Systems}
    
        ({\it Target Domain})    
    \end{multicols}
    
    \begin{multicols}{2}
        \begin{enumerate}
            \item[$P_{1}$] ML systems are complex adaptive entities.
            \begin{itemize}
                \item They seek to optimise some reward
                \item Their behaviour is implicitly specified as the result of optimising the reward (under the constraints of their architecture and computational limitations)
                \item Their actions are subject to constraints, which may be explicit or implicit
            \end{itemize}
        \end{enumerate}
    
    \columnbreak
    
        \begin{enumerate}
        \item[$P^{*}_{1}$] Economic agents are complex adaptive entities.
        \begin{itemize}
            \item They seek to optimise some reward
            \item Their behaviour is implicitly specified as the result of approximately optimising the reward while satisfying legal constraints.
            \item Their actions are subject to constrains, which may be explicit or implicit
        \end{itemize}
    \end{enumerate}
\end{multicols}

    \begin{multicols}{2}
    \begin{enumerate}
        \item[$P_{2}$] Ideal actions should be beneficial to society as a whole.
        \item[]
        \item[$P_{3}$] Alignment problem
        \begin{itemize}
            \item Bounded rationality
            \item Costly engineering/design
            \item Non-implementability
            \item Planned iteration on rewards
            \item Planned completion by third party (such as fine-tuning in light of new data / new tasks)
            \item Adversarial blinding
        \end{itemize}
        \item[]
        \item[] $\vdots$
        \item[]
        \item[$Q$] More implicit specifications work better than explicit ones for solving optimisation problems in ML systems.
    \end{enumerate}

    \columnbreak

    \begin{enumerate}
        \item[$P^{*}_{2}$] Ideal actions should be beneficial to society as a whole.
            \item[]
            \item[$P^{*}_{3}$] Principal-agent problem
            \begin{itemize}
                \item Bounded rationality
                \item Costly cognition/drafting
                \item Non-contractibility
                \item Planned renegotiation
                \item Planned completion by third party
                \item[]
                \item Strategic behaviour
            \end{itemize}
            \item[]
            \item[] $\vdots$
            \item[]
            \item[$Q*$] More implicit specifications are better than explicit ones for solving optimisation problems in economic systems.
        \end{enumerate}

    \end{multicols}
    
    \caption{Positive and hypothetical analogy between machine-learning systems and economic systems}
    \label{tab:PosAnalogy}
\end{table}

As mentioned in the introduction, our purpose is to argue for the hypothetical analogy, $Q^{*}$. Thus, our argument explicitly takes the following form: given the strength of the analogy between machine learning systems and economic systems (as specified by $P_{1}$-$P_{3}$), and given that $Q$ holds in the source domain, it is plausible that $Q^{*}$ holds in the target domain.

We suggest, following the commonsense criteria laid out by \citet{sep-reasoning-analogy}, that this inductive argument is strong. On the one hand, there are many similarities between the source and target domains. Furthermore, the conclusion, $Q^{*}$, is a relatively weak claim---it does not give particular specifications as to what the implicit rules consist in, for example. We hold that the similarities between these two systems are structural and not superficial and the similarities between these systems are relevant to our hypothetical analogy; however, \citet{sep-reasoning-analogy} notes that it is not clear why structural and causal analogies are taken as especially important, nor which structural or causal features merit attention. 

We have neither enumerated every possible positive analogy between these two systems, nor any negative analogy between them; nonetheless, our purpose to this point was to make clear {\it some} of the analogous features of these two systems to provide evidence that the analogy holds. In this sense, the conclusion of our argument---the hypothetical analogy---is {\it at least} plausible. In the next section, we discuss whether or not this conclusion is {\it practical}.

\section{Are Implicit Rules Practical?}
\label{sec:Possibility}

The success of deep learning for a broad set of AI tasks (by implicitly telling the AI what to do via an empirical objective to be optimised), compared to classical rule-based AI (for which constraints are very explicit and clearly defined), suggests that more implicit specifications work better for those tasks which humans do primarily via intuition---i.e., those belonging to what \citet{Kahneman-2011} calls {\it System 1 abilities}. Part of the reason for this discrepancy is that deep learning systems are optimised (rather than hand-crafted). Another such reason---consistent with the theoretical result outlined above from \citet{Montufar-et-al-NIPS2014}---is that the computations required for performing well at such intuitive tasks involve a mathematical function living in a very high-dimensional space (requiring a huge number of parameters). Therefore, to specify the acceptable behaviour of one of these entities using only explicit rules, one would require a massive number of rules; whereas, maximising an objective function in a rich function space, as captured by deep networks, can be done with comparatively fewer degrees of freedom.

In Section~\ref{sec:Assumptions}, we clarified the assumptions that went into our main claim: that the successes of deep learning imply that more implicit specifications work better than explicit ones for solving optimisation problems involving functions which can hardly be reduced by a manageable set of clear formal rules---i.e., the kind of function that would be required to identify morally-good conduct. In Section~\ref{sec:Analogies}, we provided the formal framework for our argument that there is an analogy between machine-learning systems and economic entities, and we specified some of the considerations that lend strength to this analogy. These two points taken together appear to imply our hypothetical analogy: that more implicit specifications should work better than explicit ones for achieving moral behaviour in economic entities---i.e., leading to their behaviour being more aligned with the common good.

Explicit rules in economic systems take the form of clearly specified laws and regulations which constrain permissible actions. Law is the enterprise of subjecting human conduct to the governance of rules. As \citet{Hadfield-2017} highlights, rules `provide the structure on which choices and interactions are built' (p. 5). Thus, the law has in place its own `meta-rules' for how the rules ought to be interpreted---this comes in the form of an authoritative figure (e.g., the supreme court, the law-speaker, a judge, etc.). The mutual recognition of an authoritative way of resolving `tough calls' to answer whether someone has violated the rules is a crucial way in which law differs from implicit rules. Humans are needed to interpret the rules precisely when they are not sufficiently clear; however, we argue that this may be a strength of such implicit rules rather than a weakness, in that it would make it more difficult for an economic agent to find loopholes in the law when the law relies more on implicit rules or principles than on formal and precise rules. The downside is that rule verification requires human brains with judgement and cannot be easily reduced to programmatic statements that can be checked mechanically.

Implicit rules might take the form of social norms and conventions, as opposed to explicit laws.\footnote{Much has been written in philosophy on social norms and conventions. Though we do not have space to do the literature justice here, see \citet{Lewis-1969, Ullmann-Margalit-1977, Schelling-1978, Coleman-1990, Vanderschraaf-1995, Bicchieri-2006,  Young-1998, Posner-1998, Posner-2000, Voss-2001, Verbeek-2002, Southwood-Eriksson-2011}.} Or, they might take the form of explicit laws which afford some degree of freedom in terms of interpretability. %
%
%Much has been written about social norms and conventions and the conceptual distinction between these two. Although, in ordinary English, these two are often used interchangeably \citep{Gilbert-1989, Gilbert-2008}, there is disagreement about the relation between norms and conventions. By way of example, \citet{Lewis-1969} suggested that conventions are special kinds of norms, which are socially enforced: `one is expected to conform, and failure to conform tends to evoke unfavourable responses from others'; others suggest that norms are special kinds of conventions.\footnote{There are several variations on this theme. See, for example, \citet{Ullmann-Margalit-1977, Young-1998, Posner-1998, Posner-2000, Coleman-1990, Voss-2001, Verbeek-2002}.} Others still suggest that there is no such conceptual or functional relation between norms and conventions \citep{Southwood-Eriksson-2011}. In general, both social conventions and social norms are taken to be endogenous products of individuals' interactions \citep{Lewis-1969, Ullmann-Margalit-1977, Schelling-1978, Vanderschraaf-1995, Bicchieri-2006}. These often emerge naturally and spontaneously, and when there is a commonly known interpretation of what the rules mean in any given situation, community enforcers may act with the knowledge that everyone else will also perceive what they are doing as within the bounds of the rules. However, they have no central authority to determine precisely what they mean.
%
In theory, laws have many significant benefits over social norms or conventions. However, in practice, these benefits are often not instantiated. For example, laws make it possible to quickly change rules, whereas social conventions often take time to `drift' via cultural evolution. In practice, though, it is often extremely difficult to change laws because of the inherent complexity of the system in which they are couched and the inefficiency or corruption of the governing bodies who can do so: \citet{Hadfield-2019} highlights, for example, that American `bar associations in the early twentieth century built a constitutional castle around the provision of legal work and established lawyers as the gatekeepers' (p. 11). 

Similarly problematic is the fact that current legal infrastructure in developed countries is often grossly outdated (in the digital era), complex, and slow to change, while in developing countries, this infrastructure is often nonexistent. The purpose of a legal framework is supposed to be to protect the freedom and autonomy of individuals (i.e., to benefit society as a whole); however, though these rules and systems of rules should be produced in a way that is of a more significant benefit to humankind, there are often not produced in such a way. Similarly, legal infrastructure often relies heavily on centralised planning, so that regulation is something that is produced solely by politicians, policymakers, and civil servants \citep[p. 246]{Hadfield-2019}.

Of course, one might worry that we are suggesting doing away with laws altogether and that this suggestion has inherently problematic consequences. The purpose of this section is to highlight the possibility of using more implicit rules relying on human judgement, by examining several real-world cases which we take to be indicative of implicit versus explicit rules {\it in practice}. These three cases are (1) distinguishing the letter and the spirit of the law, (2) shareholder theory versus stakeholder theory in a corporate legal framework, and (3) rule-based and principle-based accounting principles for economic entities.

\subsection{Example 1: The Letter and the Spirit of the Law}
\label{sec:Example1}

One way of distinguishing the letter and spirit of the law is whether expectation between parties is explicitly documented. The letter of the law---{\it litera legis} \citep{Garner-2009}---represents the explicitly documented expectations; thus, the letter of the law stands for the formal boundary between legal and illegal actions. In contrast, the spirit includes the undocumented (or tacitly held) expectations \citep{Harmon-et-al-2014}, representing the law's `general meaning or purpose, as opposed to its literal content' \citep{Garner-2009}. To put it in a slightly different way, the letter of the law is its literal meaning, whereas the spirit is its perceived intention \citep{Garcia-et-al-2014}, which requires human judgement to be identified.\footnote{There are other ways that the spirit of law is defined; for example, \citet{Ostas-2004} prioritises the fundamental rules underlying the social and ethical values protected (or supposed to be protected) by the letter of the law; \citet{Finkel-1995, Finkel-1999} understands the spirit of the law in terms of `commonsense justice', understood in terms of what the law ought to be (p. 669).} In any case, it should be clear that the spirit of the law is an inherently normative concept.

Violations of the letter of the law might be understood as the failure to fulfil a clear, documented expectation expressed in the contract; whereas, violations of the spirit of law include failures to fulfil an undocumented (though, by assumption tacitly agreed upon) expectation. \citet{Harmon-et-al-2014} discuss how parties are likely to react when contract violations occur. In particular, they find that letter violations are more difficult to overcome than spirit violations---perhaps due to perceived intentionality. 

Sometimes, breaking the letter of the law can be relatively inert. For example, though the posted speed limit on highways in Canada gives an explicit, hard rule between driving the legal speed and speeding (doing something illegal), it is a convention that it is acceptable to drive approximately 10 km per hour faster than the posted speed limit. Thus, the letter of the law on a particular stretch of the Trans-Canada Highway is `do not exceed 100 km/h'. The spirit of the law is `maintain a speed of around 100 km/h'. A Canadian may well accept being pulled over for driving $115$ km/h but may feel indignant (and perhaps justifiably-so) if pulled over for driving $102$ km/h. 

It is a point of contention in the United States whether the law ought to be interpreted on account of its letter or its spirit. For example, in American constitutional law, `originalist' or `textualist' legal scholars argue that the constitution should be interpreted by the letter of what is written since the amendment process precludes broad interpretation \citep{Easterbrook-1989, Taylor-1995}. By contrast, in Canada, the {\it Interpretation Act} of 1985 explicitly states that `the law shall be considered\ldots according to its true spirit, intent and meaning'. So, implicit rules, in the form of interpretability of spirit, are the legal standard in Canada.

\subsection{Example 2: Shareholder Theory versus Stakeholder Theory}
\label{sec:Example3}

There are laws in place to protect individuals who have a connection to an economic entity, like a corporation. Shareholders are individuals who have a vested financial interest in the economic entity, whereas stakeholders include individuals, groups, or organisations that are impacted by the outcomes of the behaviours of the economic entity. Thus, given the fact that economic entities seek to maximise something (like profits) so as to be beneficial, the question is {\it to whom should their actions benefit}. Shareholder theory suggests that economic entities are beholden solely to shareholders, whereas stakeholder theory suggests that they are also beholden to, perhaps, the wider public or society. Thus, corporations ought to consider constituents other than shareholders when making decisions to the extent that `all parties work together for a common goal and obtain shared benefits' \citep[p. 285]{Reynolds-et-al-2006}.

There is no explicit statute in the United States legally requiring corporations to maximise profits (or shareholder interests). However, case law sets a clear and unambiguous precedent---The most common reference is to {\it eBay} v. {\it Newmark} (2010), which many commentators view as a {\it decisive} affirmation that shareholder wealth maximisation is the {\it sole} legally permissible objective of a for-profit corporation \citep{Boatright-2017}. 

Section 122 of the {\it Canadian Business Corporations Act} of 1985 instead states the following:
\begin{tcolorbox}[left skip=1cm,colback=white]
\footnotesize
\singlespacing
    \begin{enumerate}
        \item[{\bf 122}] {\bf (1)} Every director and officer of a corporation in exercising their powers and discharging their duties shall
        \begin{enumerate}
        \item[]
            \item[{\bf (a)}] act honestly and in good faith with a view to the best interests of the corporation; and
            \item[{\bf (b)}] exercise the care, diligence and skill that a reasonably prudent person would exercise in comparable circumstances.
        \end{enumerate}
    \end{enumerate}
\end{tcolorbox}

Thus, it was up for interpretation what was meant by the `best interest' of the corporation. Canadian law circa the 1980s leaned toward shareholder theory; however, through the 1990s, there was a paradigm shift in case law that created some ambiguity about how to interpret this phrase. Two landmark cases set a definitive precedent for stakeholder theory in Canadian law \citep{Puri-2010, Alexander-2012}. No longer a mere precedent, on 21 June 2019 the House of Commons passed Bill C-97, which explicitly clarifies what is meant by `best interest':

\begin{tcolorbox}[left skip=1cm,colback=white]
\footnotesize
\singlespacing
\begin{enumerate}
    \item[({\bf 1.1})] When acting with a view to the best interests of the corporation under paragraph (1)(a), the directors and officers of the corporation may consider, but are not limited to, the following factors:
    \begin{enumerate}
        \item[]
        \item[{\bf (a)}] the interests of
        \begin{enumerate}
            \item[{\bf (i)}] shareholders,
            \item[{\bf (ii)}] employees,
            \item[{\bf (iii)}] retirees and pensioners,
            \item[{\bf (iv)}] creditors,
            \item[{\bf (v)}] consumers, and
            \item[{\bf (vi)}] governments;
            \item[]
        \end{enumerate}
        \item[{\bf (b)}] the environment; and
        \item[{\bf (c)}] the long-term interests of the corporation.
    \end{enumerate}
\end{enumerate}
\end{tcolorbox}
Whereas shareholder interest corresponds to profit maximisation and requires measuring only profit, which is formally quantifiable, taking into account the above stakeholders interest is hardly reduced to a simple set of verifiable or quantifiable values. Instead, they require implicit human judgement and correspond more to implicit rules or principles. 

\subsection{Example 3: Rules and Principles}
\label{sec:Example2}

The final example we will consider concerns rules surrounding generally accepted accounting principles (GAAP). This is a common set of rules, principles, standards, and procedures that economic entities must follow when compiling their financial statements. This provides a set of accounting guidelines for, e.g., revenue recognition and balance sheet classification. In an international context, this is referred to as international financial reporting standards. In the United States, GAAP is rule-based, which means that these principles can be acted on without contextualisation. In contrast, GAAP in Canada is principle-based. Similar to the letter versus the spirit of the law, this means that there is some room for judgement and flexibility in accounting so that the {\it substance} rather than just the form of a transaction is recorded. The principles, in this sense, are general by nature and require interpretation---so, their applications may vary depending on the context.

\phantom{a}

These three examples are merely meant to be illustrative of the fact that even when rules are written down explicitly, they may be open to interpretation. As such, the content of the rule is more implicit. Rather than implying that laws should be done away with altogether, this shows one way of understanding the meaning of the hypothetical analogy which we have been arguing for in this paper---that more implicit specifications are better than explicit ones for representing rich solutions to optimisation problems in economic systems. Thus, we take it that this serves as more than just a theoretical guideline, but one that is also practical and could be acted upon.

\section{Conclusion}
\label{sec:Conclusion}

We have discussed, in this paper, an analogy between the objects of study in economics and machine learning---namely, approximately rational agents. We have seen that, in trying to act rationally, both seek to optimise something. However, this optimisation process is subject to several possible constraints. The pressure exerted on these adaptive entities can come in the form of very explicit constraints or more implicit pressures. Thus, the analogy between economic entities and corporations combined with the recent successes of machine learning algorithms in performing tasks which require intuitive cognition, along with the observation that moral behaviour has not been reduced to explicit rules but can be achieved by human judgement, seem to jointly imply that more implicit specifications work better than explicit ones for capturing moral behaviour. As such, it appears that a regulatory and fiscal environment which provides more implicit pressures on corporations may lead to their behaviour being more aligned with the common good.

Here is a suggestive example deriving from this analysis. It may be more beneficial, from an economic point of view, to have more of the civil servants be allowed to exert their own judgement, like judges---i.e., by specifying their objective in a more implicit (according to high-level principles rather than very precise and explicit rules). Indeed, when trying to prescribe their behaviour by a set of clear rules (as is often the case in large organizations and bureaucracies), one obtains more consistent decisions less subject to corruption, but possibly at the price of not efficiently aligning their behaviour with the social good which is intended. This is consistent with the common observation that such rules lead to unjust or wasteful decisions. Pushing the needle towards more self-responsible civil servants along with more transparency of their decisions could improve the efficiency and moral alignment of these bureaucracies while avoiding the temptations of corruption. 

Here is another suggestive example. It may be more beneficial to modify the governing rules of for-profit corporations so that instead of maximising profit under the constraint of acting legally, they would be asked to maximise the common good, which includes being able to operate and producing useful goods and services (so selling more is good from that point of view) but also includes the impact of their actions on the rest of society (e.g. through pollution, the well-being of their employees and customers, etc.). This would require a form of auditing which includes not just financial aspects but also the evaluation of the positive and negative impact on society, which could be translated into fiscal rewards or punishments. Of course, these suggestions would make it more difficult to quantify the value brought by corporations, but human-driven systems (requiring human judgement) to do such things already exist (e.g. to evaluate how to fund non-profit organisations or research labs), so there are working precedents for more principle-based rewards for agents.

\singlespacing
\bibliographystyle{apalikelike}
\bibliography{Biblio}
\end{document}